\documentclass[11pt]{article}
\usepackage{amssymb}
\author{K.P.Tod\\Mathematical Institute\\Oxford}
\title{Isotropic Cosmological Singularities: \\Other matter models}

\begin{document}
\maketitle
\begin{abstract}
Isotropic cosmological singularities are singularities which can be
removed by rescaling the metric. In some cases already studied
(\cite{at}, \cite{a}) existence and uniqueness of cosmological models
with data at the singularity has been established. These were
cosmologies with, as source, either perfect fluids with linear
equations of state or
massless, collisionless particles. In this article we consider how to
extend these results to a variety of other matter models. These are
scalar fields, massive collisionless matter, 
the Yang-Mills plasma of
\cite{cb} and matter satisfying the Einstein-Boltzmann equation.
\end{abstract}
\section{Introduction} 
An isotropic cosmological singularity is one which can be removed by
 rescaling the spacetime metric with a single function which becomes
 singular on a smooth spacelike hypersurface, say $\Sigma$, in the
 rescaled space-time (\cite{gw},\cite{at}). In the rescaled but
 unphysical space-time the curvature
 is finite at $\Sigma$. The Weyl curvature with indices suitably
 arranged is conformally-invariant,
 so that it is also finite at the curvature singularity in the
 unrescaled physical space-time. The motivation for studying this
 class of singularities comes 
from Penrose's
Weyl Curvature Hypothesis (\cite{p1}, \cite{p2}) which can be
 interpreted 
as the hypothesis that the Weyl curvature must be finite at
 any initial 
singularity.

In order to make mathematical progress with the Weyl Curvature Hypothesis one must first say
exactly what is meant by a `finite Weyl curvature' singularity, that
is 
to say a
singularity at which the Weyl curvature is finite while, of necessity,
the Ricci curvature is not. There may
be more than one plausible way to do this. The strategy adopted 
here is first to make the
following definition (after \cite{gw}): A 
spacetime~$(\tilde{M},\tilde{g}_{ab})$~is 
said to admit an~\emph{isotropic singularity}~if there exists a
manifold~$M\supset\tilde{M}$, a regular Lorentz
metric~$g_{ab}$~on~$M$, 
and a function~$\Omega$~defined on~$M$,~such that
\begin{equation}
\tilde{g}_{ab}=\Omega^{2}g_{ab}~~~~\textrm{for}~~~\Omega>0
\label{1}
\end{equation}
\[
\Omega\rightarrow 0~~~~\textrm{on}~\Sigma
\]
where~$\Sigma$~is a smooth spacelike hypersurface in~$M$, called the
\emph{singularity surface} (note that tilded quantities are defined in
the physical 
space-time, while untilded quantities are in the unphysical
space-time; also $\Sigma$ is smooth but there need be no assumption
that $\Omega$ is smooth at $\Sigma$).
Since the Weyl tensor with its indices arranged as~$C^{a}_{~bcd}$~is 
conformally invariant and is finite in~$M$, it must also be 
finite in~$\tilde{M}$, so that isotropic singularities form a well-defined
class of cosmological singularities with finite Weyl tensor.

Consideration of the conformal Einstein equations near an isotropic 
singularity, for various matter models, leads naturally to a class of 
singular Cauchy problems for the unphysical metric~$g_{ab}$ with data
given on~$\Sigma$, for which one may seek to prove suitable existence
and uniqueness theorems. This program was begun in \cite{t1} and  \cite{t2} 
where the problem was treated in the case of a perfect 
fluid with polytropic equation of state as matter source. There the 
conformal field equations were
written as an evolution system for~$g_{ab}$ and it was shown that
solutions which are power-series in a time-coordinate could be
constructed with data 
just the
3-metric of the singularity surface. The first existence and
uniqueness theorem for the class of singular Cauchy problems which
arise was given by Claudel and Newman \cite{cn} for the particular
case when the source is a radiation fluid. This theorem was adapted to
solve the problem for other polytropic fluids in \cite{at}, and
extended to deal with the
massless Einstein-Vlasov equations, (i.e. the Einstein equations 
when the source is massless, collisionless matter satisfying the
Vlasov or Liouville equation) in \cite{a}. There
are striking differences in the freely-specifiable data in the
different cases of perfect fluid and collisionless matter. For a
perfect fluid, the initial 3-metric alone is the free data while for the
collisionless matter the free data is the initial distribution function
subject only to a `vanishing dipole' condition (the initial first and
second fundamental forms are determined by the initial distribution 
function). It is therefore a
natural question to consider other matter models, and in particular to
see how the gap between these two earlier cases may be bridged.

In this article, we shall consider a range of other matter models and
treat them in the style of \cite{t1} and \cite{t2}. That is, we shall
seek to formulate an initial value problem with data at the
singularity, to identify the data, and to show that power-series
solutions may be found. We do not seek to manipulate the problems into
the form in which the Claudel-Newman theorem can be applied as this is
generally an extremely laborious task. For certainty, this would need
to be done but in the past the study of power-series solutions has
been a 
reliable guide to existence and uniqueness in general.

The matter models we consider are three: scalar fields, a coupled
system of Einstein-Yang-Mills-Vlasov studied by Choquet-Bruhat and
collaborators \cite{cb} as a model of a `quark-gluon plasma', and
kinetic theory with either the
massive Vlasov equation or the massless 
Boltzmann equations. What we find can be summarised as follows:

For scalar fields subject to a particular condition on the growth of
the potential with large values of the field it is possible to pose an
initial value problem for which the data is just the 3-metric of the
singularity surface. There is no free data for the scalar field, which
diverges at the singularity but is asymptotically constant in spatial
directions, and the second fundamental form of the singularity surface
is automatically zero. This case is very 
much like the `stiff' perfect fluid \cite{t1}. 

The Einstein-Yang-Mills-Vlasov case is similar to the (massless) Vlasov
case \cite{a}. There are data for the initial distribution function and
the initial Yang-Mills electric field and magnetic potential. There is
a Gauss-law type constraint, and a dipole-condition connecting all the
data. There are constraints relating the first and second fundamental
forms of the singularity surface to the matter variables which seem,
and in some cases definitely are, strong enough to 
determine them uniquely in terms of the matter variables. 
The massive Vlasov case is essentially the same as the massless case:
at the singularity, rest-mass of the particles is unimportant. When we
introduce a collision term into the Vlasov (or Liouville) equation to
obtain the Boltzmann equation however there is a new phenomenon. The
collision term is defined as usual by a collision integral over
products of phase space with a kernel determining the differential
cross-section. Behaviour of the cross-section under conformal
rescaling, or equivalently in the limit of large energies, now
determines the problem. If the cross-section does not grow too fast
then the initial value problem is like the Vlasov case and the data is
the initial distribution function, which then determines the initial
metric. If the cross-section grows fast enough then there is a
constraint on the initial distribution function, that the collision
integral must vanish initially. In this case the matter is initially
in thermal equilibrium and we are back to the case of a perfect
fluid. 
The data reduces to the initial 3-metric.

The idea that a study of the Boltzmann equation would lead to an
understanding of the differences between the perfect-fluid case and
the Vlasov case is due to Alan Rendall and Keith Anguige \cite{RA}, who made
some calculations in this area, and I am grateful to them for the
suggestion.\\

The plan of the paper is as follows: we shall give a section to each of the
three models described above, after ending this section 
by reviewing our conventions for curvature and conformal rescaling.

Throughout the paper we take the spacetime metric to have
signature~$(+---)$. For a metric~$g_{ab}$~our definition of the Riemann tensor is 
\[
(\nabla_{a}\nabla_{b}-\nabla_{b}\nabla_{a})V^{c}=R^c_{~dab}V^{d}
\]
while for the Ricci tensor we take
\[
R_{ab}=R^c_{~acb}.
\]
Whenever we consider a rescaling according to (\ref{1}), tilded quantities will
always refer to the singular, physical spacetime
$(\tilde{M},\tilde{g}_{ab})$, while un-tilded quantities will refer to
the regular, unphysical spacetime 
$(M,g_{ab})$. 
For metrics $g_{ab}$ and $\tilde{g}_{ab}$ related by (\ref{1}), the Ricci tensors are related by:
\begin{eqnarray}
\tilde{R}_{ab} & = & R_{ab}- 2\nabla_{a}\nabla_{b}\log\Omega + 
2\nabla_{a}\log\Omega\nabla_{b}\log\Omega \nonumber\\
 & &- g_{ab}(\square\log\Omega
 +2\nabla_{c}\log\Omega\nabla^{c}\log\Omega) 
\label{ric2}
\end{eqnarray}
and the Weyl tensors by:
\[
\tilde{C}^{a}_{~bcd}=C^{a}_{~bcd}.
\]
If $\tilde{t}^{a}$ is a unit vector in $\tilde{M}$ then
$t^{a}=\Omega\tilde{t}^{a}$ is one in $M$, and if $\tilde{w}_{a}$ is a
unit covector in $\tilde{M}$, then $w_{a}=\Omega^{-1}\tilde{w}_{a}$ is
one in $M$.

We will usually have a cosmic time coordinate $Z$ at our
disposal. This is a smooth function in the unphysical space-time which
is constant on space-like hypersurfaces and vanishes at $\Sigma$. Its
gradient is proportional to the unit normal $N_a$ to the constant-time 
hypersurfaces:
\begin{equation}
Z_a=\nabla_aZ=VN_a
\label{zzz}
\end{equation}
where $V^2=g^{ab}Z_aZ_b$. 

The intrinsic metric $h_{ab}$ on the surfaces of constant $Z$, which is also the projection orthogonal to $N^{a}$, is defined by
\begin{equation}
h_{ab}=g_{ab}-N_{a}N_{b}
\label{met1}
\end{equation}
so that $h_{ab}$ is negative-definite. We shall use $D_a$ for the intrinsic 3-dimensional metric covariant derivative on surfaces of constant $Z$.

The second-derivative of $Z$ defines the second fundamental form $K_{ab}$ of the constant-time hypersurfaces by the equation
\begin{equation}
\nabla_a\nabla_bZ=V(K_{ab}+N_aA_b+A_aN_b+V_ZN_aN_b)
\label{K}
\end{equation}
where $A_a=h_a^{\,b}\nabla_b\log V = D_a\log V$ is the acceleration of the normal congruence defined by $N_a$, and $V_Z=V^{-1}N^a\nabla_aV$ (since in general $N^a\nabla_a~=~V\partial/\partial Z$).

For the Cauchy problem with data on surfaces of constant $Z$ various identities relating 3 and 4-dimensional curvatures are useful. These are
\begin{eqnarray}
G_{ab}N^aN^b&=&-\frac{1}{2}({}^3R-K^2+K^{ab}K_{ab})
\label{ham}\\
R_{ac}N^ah_b^{\,c}&=&D_cK_b^{\,c}-D_bK
\label{mom}\\
R_{cd}h_a^{\,c}h_b^{\,d}&=& -{\mathcal{L}}_N K_{ab}
+{}^3R_{ab}+2K_{ac}K_b^{\,c}-KK_{ab}\nonumber\\
&&+D_aA_b-A_aA_b
\label{ev}
\end{eqnarray}
where $G_{ab},R_{ab}$ are the 4-dimensional Einstein and Ricci tensors,
$^3R, {}^3R_{ab}$ are the 3-dimensional Ricci scalar and tensor,
$K=h^{ab}K_{ab}$ is the trace of the extrinsic curvature, and
${\mathcal{L}}_N$ is the Lie-derivative along $N^a$. When we impose
the Einstein equations, these identities give the Hamiltonian and
momentum constraints and the evolution equation for 
$K_{ab}$, respectively.

The field equations in \cite{a} and \cite{at} were obtained in a
first-order form as:
\begin{equation}
A^0(X)\frac{\partial X}{\partial Z} = A^i(X)\frac{\partial X}{\partial x^i}
+\frac{1}{Z}B(X)X+C(X,Z).
\label{fuch}
\end{equation}
Here $X$ is a multi-component vector of unknowns including the first
and second fundamental forms of the constant-$Z$ surfaces and matter
variables, $Z$ is the cosmic time coordinate, $x^i$ are comoving
spatial coordinates, $A^0$ and $A^i$ are symmetric matrices with $A^0$
positive definite, $C$ is analytic in $Z$ and $B$ is subject to
conditions which will emerge. There is a singularity in the time at $Z=0$ but,
by assumption, nowhere else.

Suppose we are given data $X_0$ at $Z=0$. If we
want (\ref{fuch}) to hold at $Z=0$ we at once have a condition, that
the term $B(X_0)X_0$ must vanish. We call this the Fuchsian condition
for this system. If we now substitute a power series in positive
powers of $Z$ we can determine the coefficients of the series provided
none of the eigenvalues of $(A^0)^{-1}B(X)$ is a positive
integer. With essentially these assumptions, Claudel-Newman \cite{cn} prove
existence and uniqueness of solution.

Putting field equations into precisely this first-order form is
therefore the acid test of existence and uniqueness. However, this is
difficult to do. One can recognise the Fuchsian conditions and
impose the eigenvalue condition without going that far and this is
what we shall do below.

\section{Scalar fields}
We begin by considering cosmological models whose source is a scalar
field. We shall find that these may have isotropic singularities
provided the potential satisfies a condition and that there is then an
initial value problem for which the data is the 3-metric of the
singularity surface $\Sigma$. As one might expect, this case is
similar to the perfect fluid with pressure equal to 
density (see \cite{t1}).

If the source of the gravitational field is a scalar field $\phi$ with potential $U(\phi)$ then the energy-momentum tensor in the physical space-time is
\[
\tilde{T}_{ab}=\phi_a\phi_b-\tilde{g}_{ab}(\frac{1}{2}\tilde{g}^{cd}\phi_c\phi_d+U(\phi))
\]
writing $\phi_a$ for $\nabla_a \phi$. The conservation 
equation for this $\tilde{T}_{ab}$ is satisfied by virtue of the 
field equation for the scalar field which is
\begin{equation}
\tilde{\square}\phi=U'(\phi).
\label{phi}
\end{equation}
The tensor which we want to substitute into (\ref{ric2}) is the trace-reversed energy-momentum tensor:
\begin{equation}
\tilde{T}_{ab}-\frac{1}{2}\tilde{T}\tilde{g}_{ab}
=\phi_a\phi_b+\tilde{g}_{ab}U(\phi).
\label{ems2}
\end{equation}
Now from (\ref{ric2}), (\ref{ems2}) and the Einstein equations we find
\begin{eqnarray}
R_{ab}&=& \frac{2}{\Omega}\nabla_a\nabla_b\Omega -\frac{4}{\Omega^2}\nabla_{a}\Omega\nabla_{b}\Omega 
 + g_{ab}(\frac{1}{\Omega}\square\Omega +\frac{1}{\Omega^2}\nabla_{c}\Omega\nabla^{c}\Omega)\nonumber\\
&&+\frac{8\pi G}{c^2}(\phi_a\phi_b+\Omega^2g_{ab}U(\phi)).
\label{ric3}
\end{eqnarray}
where we don't yet know what to take for $\Omega$, but if there is to
be an isotropic cosmological singularity then this must hold for some
$\Omega$, and $\tilde{T}_{ab}$ must be singular at $\Omega=0$. 
In particular $R_{ab}$ is finite at $\Sigma$, where $\Omega$
vanishes, and by (\ref{ric3}) at $O(\Omega^{-2})$ this requires
\begin{equation}
\phi_a\sim\frac{\Omega_a}{\Omega}
\label{phi2}
\end{equation}
in this limit. Looking carefully at (\ref{ric3}), we see that we cannot require $\Omega$ to be smooth at $\Sigma$ but we can require $Z=\Omega^2$ to be smooth (just as for the perfect fluid with $p=\rho$). Now to satisfy (\ref{phi2}) we need $\phi$ to be a function of $Z$ at least as $Z$ tends to zero and $\phi$ to infinity. Thus near the singularity, $\phi$ tends to infinity and the surfaces of constant $\phi$ are space-like. Consequently we can take a function of $\phi$ to be the cosmic time coordinate $Z$ and use this to define $\Omega$. The correct function to take is determined by (\ref{ric3}) as
\begin{equation}
Z=\Omega^2=\exp(\frac{\phi}{\alpha})
\label{z2}
\end{equation}
where $\alpha$ is a constant to be determined (if $\alpha>0$ then 
$\phi\rightarrow-\infty$ as $Z\rightarrow 0$ while if $\alpha<0$ 
then $\phi\rightarrow+\infty$). We choose units with 
$\frac{8\pi G}{c^2}~=~1$ and substitute (\ref{z2}) into (\ref{ric3}) to find
\begin{equation}
ZR_{ab}= \nabla_a\nabla_b Z 
-\frac{1}{Z}(\alpha^2-\frac{3}{2})\nabla_a Z\nabla_b Z 
 + (\frac{1}{2}\square Z  + Z^2U(\phi)) g_{ab}
\label{ric4}
\end{equation}
where we have also multiplied through by $Z$. Now we can see that if
(\ref{ric4}) is to hold at $\Sigma$, where $Z=0$, then necessarily
$\alpha^2=3/2$. Furthermore, with this choice we have a regular field
equation in the unphysical space-time provided the term with the
potential $U$ is finite at $\Sigma$. This requires $Z^2U(\phi)$ to be
finite as $Z$ tends to zero, or equivalently
$\exp(\frac{2\phi}{\alpha})U(\phi)$ to be finite as $\phi \rightarrow
\pm\infty$ 
where the correct choice of sign depends on the sign of $\alpha$.

If we calculate the contracted Bianchi identity in the unphysical space-time using (\ref{ric4}) then we find that this is equivalent to the equation
\begin{equation}
\square Z=\frac{1}{\alpha}Z^2U'(\phi)
\label{z3}
\end{equation}
which can also be obtained from (\ref{phi}) by conformally transforming and changing the dependent variable.

We next need to formulate a Cauchy problem for the Einstein equations in the form (\ref{ric4}) with the matter equation (\ref{z3}). We perform a 3+1-splitting with respect to the hypersurfaces of constant $Z$ and use $Z$ as a time-coordinate. The variables in the Cauchy problem will be $V, h_{ab}$ and $K_{ab}$ at each value of $Z$ and we need evolution equations for each of these, with the possibility of constraints between them.

In terms of comoving coordinates $x^i, i=1,2,3$ the metric can be written as
\begin{equation}
ds^2=\frac{dZ^2}{V^2}+h_{ij}dx^idx^j
\label{met2}
\end{equation}
with $V$ and $h_{ij}$ as in (\ref{zzz}) and (\ref{met1}).

From the trace of (\ref{K}) we obtain
\begin{equation}
\square Z=V(K+\frac{\partial V}{\partial Z})
\label{z4}
\end{equation}
where $K=g^{ab}K_{ab}$, which with (\ref{z3}) gives an evolution equation for $V$. Substituting from (\ref{K}) and (\ref{z3}) into (\ref{ric4}) puts the Einstein equations into the form
\begin{equation}
ZR_{ab}=V(K_{ab}+N_aA_b+A_aN_b+V_ZN_aN_b)
+g_{ab}(\frac{1}{2\alpha}Z^2U'(\phi)+Z^2U(\phi)).
\label{ric5}
\end{equation}
From (\ref{ric5}) and (\ref{ham}) we obtain the Hamiltonian constraint as
\begin{equation}
-2G_{ab}N^aN^b= {}^3R-K^2+K_{ab}K^{ab}=2(\frac{VK}{Z}+ZU(\phi))
\label{ham1}
\end{equation}
and from (\ref{mom}) the momentum constraint as
\begin{equation}
R_{ac}N^ah_b^{\,c}=D_cK_b^{\,c}-D_bK=\frac{1}{Z}D_b\log V
\label{mom1}
\end{equation}
where we have used the definition of $A_a$ from (\ref{K}). For the evolution equations we have the evolution of $V$ from (\ref{z3}) and (\ref{z4}):
\begin{equation}
\frac{\partial V}{\partial Z}=-K+\frac{1}{\alpha V}Z^2U'(\phi)
\label{vev}
\end{equation}
then the evolution of $h_{ab}$, which is essentially the definition of $K_{ab}$:
\begin{equation}
{\mathcal{L}}_Nh_{ab}=2K_{ab}
\label{hev}
\end{equation}
and finally from (\ref{ev}) and (\ref{ric5}) the evolution of $K_{ab}$
\begin{eqnarray}
{\mathcal{L}}_N K_{ab}&=&
^3R_{ab}+2K_{ac}K_b^{\,c}-KK_{ab}+D_aA_b-A_aA_b -\frac{V}{Z}K_{ab}\nonumber\\ 
&&+h_{ab}(\frac{1}{2\alpha}Z^2U'(\phi)+Z^2U(\phi)).
\label{kev}
\end{eqnarray}
To summarise, the evolution equations are (\ref{vev}), (\ref{hev}) and
(\ref{kev}), with constraints (\ref{ham1}) and (\ref{mom1}). Because
the contracted Bianchi identity for (\ref{ric4}) is (\ref{z3}) which
in turn is (\ref{vev}) and part of the system, we know that the
evolution preserves the constraints. The evolution equations have a
singularity in the `time' at $Z=0$. In order to put them in the form
of (\ref{fuch}) it would be necessary to introduce many more variables to make
the system first-order in space. Rather than do this, we can draw
similar conclusions in the present form. Specifically we can read off
the Fuchsian constraints and check the eigenvalue conditions noted
after (\ref{fuch}).

 For the Fuchsian constraints, if (\ref{mom1}) is to hold at $Z=0$ then we
need $D_bV=0$ so that $V$ is constant on $\Sigma$. By a constant
rescaling we can suppose that $V=1$ there. Now 
for (\ref{ham1}) at $Z=0$ we need initially
\begin{equation}
K=-\lim(Z^2U(\phi))
\label{k2}
\end{equation}
while from (\ref{kev}) we need initially
\begin{equation}
K_{ab}=-h_{ab}(\lim(\frac{1}{2\alpha}Z^2U'(\phi)+Z^2U(\phi)).
\label{k3}
\end{equation}
For consistency between (\ref{k2}) and (\ref{k3}) we need to impose
\begin{equation}
Z^2U(\phi)+\frac{3}{4\alpha}Z^2U'(\phi)\rightarrow 0\,\,\,as\,\,Z\rightarrow 0.
\label{phi3}
\end{equation}
This is in addition to the condition found earlier that $Z^2U(\phi)$
have a finite limit as $Z\rightarrow 0$. Assuming this condition on
$U(\phi)$ holds then the Fuchsian conditions can be solved to find
that the free data at $\Sigma$ consist of just the initial 3-metric
$h_{ab}$. The other variables are $V$, which is one without loss of
generality, and $K_{ab}$, which is determined by (\ref{k3}). 

It is
straightforward to construct solutions as power series in $Z$, if one
assumes power series in $Z$ for $U$, and these are unique with the
given data (this happens because the only singular evolution equation
is (\ref{kev}) and the singular term has a negative coefficient). 
While we don't convert the system into a form in which the
theorem of Claudel-Newman \cite{cn} can be applied, the existence of
unique power series solutions has in the past been a reliable guide to
the applicability of this 
theorem.

It is worth noting that $\Sigma$ is necessarily umbilic, in that its
second fundamental form is a multiple of its metric, and this multiple
is in fact a constant. Consequently the magnetic part of the Weyl
tensor (see e.g. \cite{gw}) is zero at $\Sigma$ and the electric part
of the Weyl tensor is proportional to the trace-free part of the
intrinsic Ricci tensor of $\Sigma$. The initial Weyl tensor is finite,
as it must be, and is zero if and only if $\Sigma$ has vanishing
trace-free Ricci tensor or equivalently has a homogeneous and
isotropic metric. But now uniqueness for the Cauchy problem would mean
that the cosmological model was a Friedman-Robertson-Walker
model. Thus with scalar fields we have the same situation as with
polytropic perfect fluids, namely that if the 
Weyl tensor is zero initially then it is always zero.

\section{The Einstein-Yang-Mills-Liouville equations}
The Einstein-Yang-Mills-Liouville equations have been considered by Choquet-Bruhat and coworkers as a model of the (early) universe where the matter content is a `quark-gluon plasma' \cite{cb}. The model has a distribution function for massless, collisionless matter with colour-charges which produces a Yang-Mills field. With space-time indices $a,b,...$ and Lie algebra indices $\alpha, \beta,...$ the matter variables are the Yang-Mills potential $\tilde{A}_a^{\,\alpha}$ and the distribution function $\tilde{f}(x^a,\tilde{p}_b,q^{\alpha})$ where the vector $q^\alpha$ is a vector of colour-charges. 

We retain the convention that tilded quantities relate to the physical space-time and untilded to the rescaled, unphysical space-time. In fact $A_a^{\,\alpha}, f$ and $p_b$ are all unchanged under rescaling, as is the Yang-Mills field $F_{ab}^{\alpha}$, which is defined by
\begin{equation}
F_{ab}^{\alpha}= \nabla_aA_b^{\,\alpha}-\nabla_bA_a^{\,\alpha} + c^{\alpha}_{\,\,\beta \gamma}A_a^{\,\beta}A_b^{\,\gamma}
\label{F1}
\end{equation}
where $c^{\alpha}_{\,\,\beta \gamma}$ are the structure constants for the Lie algebra, say ${\mathcal{G}}$.

The constituent `particles' of the collisionless matter follow the
 Yang-Mills version of the Lorentz force law which, for a particle
 with colour-charges given by
 $q^{\alpha}$ can be written as the system:
\begin{eqnarray}
\frac{dx^a}{ds} & = & g^{ab}p_b\nonumber\\
\frac{dp_a}{ds} & = & 
-\frac{1}{2}\frac{\partial g^{bc}}{\partial x^a}p_bp_c +q^{\alpha}\eta_{\alpha\beta}F_{ab}^{\beta}p^b\label{xdot}\\
\frac{dq^{\alpha}}{ds} & = & - c^{\alpha}_{\,\,\beta \gamma}A_a^{\,\beta}p^aq^{\gamma}\nonumber
\end{eqnarray}
here $\eta_{\alpha \beta}$ is the metric on ${\mathcal{G}}$ (assuming it is semi-simple). Again the system (\ref{xdot}) is conformally-covariant in that it takes the same appearance if tildes are inserted throughout.

The Vlasov or Liouville equation for the distribution function follows from (\ref{xdot}) and the chain rule:
\[
\frac{\partial f}{\partial x^a} \frac{dx^a}{ds}+
\frac{\partial f}{\partial p_a}\frac{dp_a}{ds}+
\frac{\partial f}{\partial q^\alpha}\frac{dq^\alpha}{ds}=0.
\]
The Yang-Mills current is obtained as an integral over the distribution function. Now we do need to watch the conformal rescaling. In the unphysical space-time we consider the vector field:
\[
J_a^{\,\alpha}=\int q^\alpha p_a f \omega_p\omega_q
\]
where $\omega_p$ is the standard Lorentz-invariant volume-form on the null cone, which can be written with respect to a 3+1-splitting as
\begin{equation}
\omega_p=\frac{1}{p^0\sqrt{-g}}dp_1dp_2dp_3
\label{vol}
\end{equation}
and $\omega_q$ is an invariant volume-form on ${\mathcal{G}}$.

Under conformal rescaling we have
\begin{equation}
\tilde{\omega}_p = \Omega^{-2}\omega_p
\label{vol2}
\end{equation}
and then the physical Yang-Mills current is
\begin{equation}
\tilde{J}_a^{\,\alpha}  =  \Omega^{-2}J_a^{\,\alpha}.
\label{j2}
\end{equation}
The physical Yang-Mills equation relates the divergence of the physical Yang-Mills field to the physical Yang-Mills current. Now this equation transforms well under conformal rescaling and in the unphysical variables we find
\begin{equation}
\nabla_aF^{ab\alpha}+ 
c^{\alpha}_{\,\,\beta \gamma}A_a^{\,\beta}F^{ab\gamma}
=J^{b\alpha}
\label{ym1}
\end{equation}
The energy-momentum tensor is a sum of two terms. For the Yang-Mills field consider
\[
T_{ab}^{YM}=\eta_{\alpha\beta}(F_{ac}^{\alpha} F_b^{\beta c}-
\frac{1}{4}g_{ab}F_{cd}^{\alpha}F^{cd\beta})
\]
and for the collisionless matter consider
\[
T_{ab}^{L}=\int fp_ap_b\omega_p\omega_q.
\]
Note that both these tensors are trace-free. The physical energy-momentum tensor is then
\begin{equation}
\tilde{T}_{ab}=\Omega^{-2}(T_{ab}^{YM}+T_{ab}^{L}).
\label{em2}
\end{equation}
We write down the Einstein equations in the unphysical space-time with
the aid of (\ref{ric2}). Following the argument in \cite{a} we may
choose the cosmic time coordinate $Z$ so that $\square Z=0$ everywhere
and $V=1$ at $\Sigma$ and we can then take $Z=\Omega$ (there is
sufficient conformal gauge freedom to allow this, provided there is an
isotropic cosmological singularity, because the energy-momentum tensor
is trace-free). From (\ref{ric2}) 
we obtain
\begin{equation}
Z^2R_{ab}=2Z\nabla_a\nabla_bZ-4\nabla_aZ\nabla_bZ +V^2g_{ab}
+T_{ab}^{YM}+T_{ab}^{L}.
\label{ric6}
\end{equation}
The dependent variables on each surface of constant $Z$ are $V, h_{ab}$ and $K_{ab}$ as in section 2, together with the Yang-Mills potential $A_a^{\,\alpha}$ and the Yang-Mills electric field $E_a^{\,\alpha}~:=~F_{ab}^\alpha N^b$, and the distribution function $f$. There is sufficient freedom in the conformal gauge to ensure that $K=0$ at $\Sigma$. We choose the temporal Yang-Mills gauge so that $N^aA_a^{\,\alpha}=0$. The Yang-Mills equations (\ref{ym1}) imply a Gauss-law constraint:
\[N_b(J^{b\alpha}-\nabla_aF^{ab\alpha}- 
c^{\alpha}_{\,\,\beta \gamma}A_a^{\,\beta}F^{ab\gamma})=0\]
which translates as
\begin{equation}
D_aE^{a\alpha}+ c^{\alpha}_{\,\,\beta \gamma}A_a^{\,\beta}E^{a\gamma} 
=\frac{1}{\sqrt {-h}}\int q^\alpha fdp_1dp_2dp_3.
\label{gc1}
\end{equation}
This constraint holds on every surface of constant $Z$ and in particular needs to hold on the data at $\Sigma$.

We perform a 3+1-splitting with respect to the hypersurfaces of
constant $Z$ as before, and decompose the Einstein equations
(\ref{ric6}) into evolution and constraints. The Fuchsian constraints
for data at $\Sigma$ will follow from the condition that the
right-hand-side in (\ref{ric6}) must be $O(Z^2)$. We write spatial
indices $i,j,k, \dots =1,2,3$ and then the conditions arise from the
vanishing of the $O(1)$ terms in the $(i,j)$ and $(0,i)$ components in
(\ref{ric6}) and the vanishing of the $O(Z)$ terms in the $(i,j)$
components. (The $O(Z)$ terms in the $(0,i)$ components vanish by
virtue of these conditions since the contracted Bianchi conditions
hold.) The conditions are most easily written in terms of 
multipole moments $\chi_i, \chi_{ij}, \chi_{ijk}, \dots$ of the matter density and $J_i^{\alpha}, J_{ij}^{\alpha}, J_{ijk}^{\alpha}, \dots$ of the charge density which we define at $\Sigma$ by
\begin{eqnarray*}
\chi_i & = & \frac{1}{\sqrt{-h}} \int f p_i d^3p\omega_q\label{c1}\\
\chi_{ij} & = & \frac{1}{\sqrt{-h}} \int \frac{f p_ip_j}{p^0} d^3p\omega_q\label{c2}\\
\chi_{ijk} & = & \frac{1}{\sqrt{-h}} \int \frac{f p_ip_jp_k}{(p^0)^2} d^3p\omega_q \label{c3}
\end{eqnarray*}
and so on, and
\begin{eqnarray*}
J_i^{\alpha} & = & \frac{1}{\sqrt{-h}} \int \frac{q^{\alpha}f p_i}{p^0} d^3p\omega_q\label{m1}\\
J_{ij}^{\alpha} & = &\frac{1}{\sqrt{-h}} \int \frac{q^{\alpha}f p_ip_j}{(p^0)^2} d^3p\omega_q\label{m2}\\
J_{ijk}^{\alpha} & = & \frac{1}{\sqrt{-h}} \int \frac{q^{\alpha}f p_ip_jp_k}{(p^0)^3} d^3p\omega_q \label{m3}
\end{eqnarray*}
and so on, where it is understood that $f,h$ and $p^0$ are evaluated at $\Sigma$. We have used (\ref{vol}) and in these definitions 
\[(p^0)^2=-h^{ij}p_ip_j.\]

Now the Fuchsian conditions are
\begin{eqnarray}
0&=&h_{ij}+\chi_{ij} -\eta_{\alpha \beta}(E_i^{\alpha}E_j^{\beta}-
F_{ik}^{\alpha}F_j^{k\beta}\nonumber\\
&& +\frac{1}{4}h_{ij}(2E_k^{\alpha}E^{k\beta}
+F_{km}^{\alpha}F^{km\beta}))
\label{f1}\\
0&=&\chi_i- \eta_{\alpha\beta}E^{j\alpha}F_{ij}^{\beta}
\label{f2}
\end{eqnarray}
and the most complicated which can be written
\begin{eqnarray}
K^{lm}Q_{lmij}&=&h^{lm}D_l\chi_{ijm}+ 
c^{\alpha}_{\,\beta\alpha}A^{\beta k}\chi_{ijk}\nonumber\\
&&-\eta_{\alpha \beta}(4E_{(i}^{\alpha}E_{j)}^{\beta}
+2F_{(i}^{m\alpha}J_{j)m}^{\beta}
+E^{k\alpha}J_{ijk}^{\beta}\nonumber\\
&&-\tilde{D}_kE_{(i}^{\alpha}F_{j)}^{k\beta}
-\tilde{D}_{(i}E_{|k|}^{\alpha}F_{j)}^{k\beta}
+2E_{(i}^{\alpha}\tilde{D}^kF_{j)k}^{\beta}\nonumber\\
&&+h_{ij}( \tilde{D}^k (F_{km\alpha}E_m^{\beta})-
E^{\alpha k}J_k^{\beta}))
\label{f3}
\end{eqnarray}
where
\begin{eqnarray}
Q_{lmij}& =& 4h_{li}h_{mj} -h_{lm}h_{ij} -\chi_{lmij}\nonumber\\
&&-
\eta_{\alpha \beta}(E_l^{\alpha}E_i^{\beta}h_{mj}
+E_l^{\alpha}E_j^{\beta}h_{mi} + E_m^{\alpha}E_i^{\beta}h_{lj} 
+ E_m^{\alpha}E_j^{\beta}h_{li} \nonumber\\
&&- E_i^{\alpha}E_j^{\beta}h_{lm} -E_l^{\alpha}E_m^{\beta}h_{ij}\nonumber\\
&&-F_{il}^{\alpha}F_{jm}^{\beta}-F_{im}^{\alpha}F_{jl}^{\beta}
+h_{ij}F_{lk}^{\alpha}F_m^{k\beta}
+h_{lm}F_{ik}^{\alpha}F_j^{k\beta}\nonumber\\
&&-h_{li}h_{mj}(2E_k^{\alpha}E^{k\beta}
+F_{kn}^{\alpha}F^{kn\beta})\nonumber\\
&&+h_{lm}h_{ij}(\frac{1}{2}E_k^{\alpha}E^{k\beta} 
- \frac{1}{4}F_{kn}^{\alpha}F^{kn\beta})).
\label{f4}
\end{eqnarray}
If we consider first (\ref{f1}), then we recall from \cite{at} that in
the case of vanishing Yang-Mills fields this actually determines the initial
metric $h_{ij}$ from the initial distribution function $f$. It is hard
to see whether this remains true for arbitrary Yang-Mills fields but
it will certainly remain true 
for small fields. 

Next note that, by virtue of (\ref{f1}) and (\ref{f2}), $Q_{lmij}$ is
symmetric and trace-free on each index pair, and symmetric under
interchange of the two pairs. If we think of it as a map from
trace-free symmetric tensors to trace-free symmetric tensors, it
necessarily has real eigenvalues and we can solve (\ref{f3}) for
$K_{ij}$ provided none of these eigenvalues is zero (recall that
$K_{ij}$ is trace-free at $\Sigma$ by choice of conformal
gauge). Again this is a difficult question in general although we know
from \cite{at} that it is true if the Yang-Mills fields are zero, so
that it will be true if they are 
small.

Finally (\ref{f2}) and (\ref{gc1}) are constraints on the initial
matter variables. After the Fuchsian conditions we should check the
eigenvalues as in the discussion following (\ref{fuch}), but again we
can argue that they must still satisfy the required condition, at
least for small fields.

To summarise, the data at $\Sigma$ consists of the Yang-Mills
 potential $A_i^{\alpha}$, from which the magnetic field 
$F_{ij}^{\alpha}$ is calculated by (\ref{F1}), the Yang-Mills 
electric field $E_i^{\alpha}$, the distribution function $f$ and 
the first and second fundamental forms $h_{ij}$ and $K_{ij}$. 
These are subject to the Gauss-law constraint (\ref{gc1}) and 
the three Fuchsian constraints (\ref{f1}), (\ref{f2}) and (\ref{f3}). 
In the pure Vlasov case, with the Yang-Mills fields omitted, 
(\ref{f1}) suffices to determine $h_{ij}$ from $f$ and (\ref{f3})
 then determines $K_{ij}$ from $f$ (see \cite{at} or \cite{a} for 
details of this). Thus in that case the free data is just $f$ subject 
to the counterpart of (\ref{f2}). With the Yang-Mills fields present 
it is not so easy to see that (\ref{f1}) determines a unique $h_{ij}$ 
for arbitrary Yang-Mills fields and that (\ref{f3}) determines a 
unique $K_{ij}$, but this must certainly be true for small fields 
by continuity.
Once the Fuchsian constraints are satisfied, and relying on the 
experience of earlier cases, we expect to have existence and 
uniqueness of solutions.

\section{Massive Einstein-Vlasov and Einstein-Boltzmann}
In this section we do two things. First we look at massive
Einstein-Vlasov in the expectation that near the singularity there
should be no difference between massless collisionless matter and
massive collisionless matter. Then, restricting to massless for
simplicity we seek to drop the collisionless condition. This is partly
to move to a more realistic situation but also to resolve a 
puzzle left over from earlier work and noted in the Introduction.

For an isotropic singularity with a perfect fluid (at least for linear
equations of state \cite{at}) the data are just the initial
3-metric. There are no separate data for the matter. For an isotropic
singularity with massless collisionless matter the data is just the
initial distribution function, which determines the initial first and
second fundamental forms \cite{at}. This time there are no separate
data for the geometry. It was suggested to me by Alan Rendall \cite{RA} that by
including a collision term in the Vlasov equation, in other words by
turning to the Einstein-Boltzmann equations, one might find a bridge
between 
these two extreme examples. This turns out to be the case.

We begin by considering the massive Einstein-Vlasov equations, by
which we mean the Einstein equations whose source is collisionless
matter where the matter is now supposed to be composed of particles of
a single nonzero mass, say $m$. The presence of the mass changes the
detail of the equations but we shall 
see that that change has no influence near to the singularity.

In this case, the stress-energy-momentum tensor in the physical variables is 
\begin{equation}
\tilde{T}_{ab}^{L}=\int fp_ap_b\tilde{\omega}_p
\label{vlas3}
\end{equation}
with
\begin{equation}
\tilde{\omega}_p=\frac{1}{\tilde{p}^0\sqrt{-\tilde{g}}}dp_1dp_2dp_3.
\label{vol3}
\end{equation}
The distribution function $f$ is supported on the mass shell where
\[\tilde{g}^{ab}p_ap_b=m^2\]
which we must solve to find $\tilde{p}^0$ in (\ref{vol3}). We
subsequently regard $\tilde{p}^0$ (and later $p^0$) as a function
of $p_i$ and the coordinates, and the distribution function as a
function only of $p_i$ and 
$x^a$. 

Under conformal rescaling according to (\ref{1}) we still have (\ref{vol2}) with $\omega_p$ as in (\ref{vol}) but now with
\begin{equation}
p^0=V^2p_0=V(-h^{ij}p_ip_j+m^2\Omega^2)^{\frac{1}{2}}.
\label{p01}
\end{equation}
As one might expect, the mass becomes insignificant near the
singularity, where $\Omega$ vanishes. The unphysical energy-momentum
tensor to go into 
(\ref{ric6}) may be written, with the aid of the metric (\ref{met2}) 
and the expression (\ref{p01}) as
\[
T_{ab}^{L}=\frac{1}{\sqrt{-h}}\int 
\frac{fp_ap_bdp_1dp_2dp_3}{(-h^{km}p_kp_m+m^2\Omega^2)^{\frac{1}{2}}}.
\]
Provided $f$ is supported away from the origin in $p$-space, which one
usually assumes for regularity (see e.g. \cite{a}), the energy
momentum tensor can be expanded as a series in positive powers of
$\Omega$. In particular in the limit at $\Sigma$ it is the same as in
the massless case. Consequently in 
rescaling we take $\Omega$ to be $Z$ just as in the massless case.

We also have to change the Vlasov equation in the presence of
mass. The Vlasov equation expresses the vanishing of the derivative of
the distribution function along the geodesic flow, or equivalently the
vanishing of its Poisson 
bracket with the Hamiltonian. In the unphysical variables this is
\begin{eqnarray}
{\mathcal{L}}_{\tilde X}f(p_a,x^b) & = & \{\frac{1}{2}\tilde{g}^{ab}p_ap_b,f\}_{PB}\nonumber\\ 
& = & \tilde{g}^{ab}p_a\frac{\partial f}{\partial x^b}
-\frac{1}{2}\frac{\partial\tilde{g}^{ab}}{\partial
x^c}p_ap_b\frac{\partial f}{\partial p_c}\nonumber\\
& = & 0.
\label{vlas4}
\end{eqnarray}
Written in this form, it is easy to see how to conformally transform the equation. With the metric as in (\ref{met2}) and rescaling as in (\ref{1}) we obtain
\begin{equation}
V^2p_0\frac{\partial f}{\partial Z}+
h^{ij}p_i\frac{\partial f}{\partial x^j} 
-(V\frac{\partial V}{\partial x^i}(p_0)^2+\frac{1}{2}
\frac{\partial h^{jk}}{\partial x^i}p_jp_k)\frac{\partial f}{\partial p_i}
+m^2\Omega\frac{\partial\Omega}{\partial x^i}\frac{\partial f}{\partial p_i} =0.
\label{vlas5}
\end{equation}
The last term in this actually vanishes because $\Omega$ is $Z$, but there are still explicit appearances of $Z$ in (\ref{vlas5}) through $p_0$. However these are always in the numerator and we get no new singularities from them.

We conclude that, at least at the level of looking for power series solutions in the time-variable, there should be no difference between massive and massless Vlasov.

To go from Vlasov to Boltzmann is to go from collisionless matter to the inclusion of a collision-term in (\ref{vlas4}). Symbolically we have
\begin{equation}
{\mathcal{L}}_{\tilde{X}}f=C(f,f)
\label{B1}
\end{equation}
where $C(f,f)$ is a bilinear functional in $f$. We set up the
formalism for this following Choquet-Bruhat \cite{cb}. We restrict 
to 2-body collisions and suppose that incoming particles with 
4-momenta $p^{(0)}$ and $p^{(1)}$ (temporarily suppressing space-time indices)
 collide to give outgoing 
particles with 4-momenta $p^{(2)}$ and $p^{(3)}$, where all 
particles have mass $m$ and total 4-momentum $P$ is conserved:
\[P=p^{(0)}+p^{(1)}=p^{(2)}+p^{(3)}\]
If we choose $p^{(0)}$ then there are 3 degrees of freedom in
$p^{(1)}$, two in $p^{(2)}$ and none in $p^{(3)}$. The collision 
term is written as an integral:
\begin{equation}
C(f,f)=\int (f_2f_3-f_0f_1) 
k(x^a,p^{(0)},p^{(1)},p^{(2)},p^{(3)})\tilde{\omega}_{p^{(1)}}
\tilde{\xi}_2
\label{B2}
\end{equation}
where $f_n=f(x^a,p^{(n)}_i)$, $k$ is a kernel to be discussed 
below, $\tilde{\omega}_p$ is as in (\ref{vol}) but with the 
time-direction defined by $p^{(0)}$, and $\tilde{\xi}_2$ is an 
invariant (Leray) 2-form on the space of allowed $p^{(2)}$ 
(see \cite{cb} for a precise definition).
The kernel $k$ is related to the differential cross-section and 
is usually assumed, for reasons of symmetry, to be a function only 
of the total energy involved in the collision, which is
\begin{equation}
\tilde{s}=\tilde{g}^{ab}P_aP_b
\label{s}
\end{equation}
and the angle $\theta$ between $p^{(0)}-p^{(1)}$ and 
$p^{(2)}-p^{(3)}$, which is the scattering angle in the 
centre-of-mass frame, so we may write $k(\tilde{s},\theta)$.
For Choquet-Bruhat \cite{cb} $k$ is the differential cross-section 
but for other authors, e.g. Grout et al \cite{glw}, 
$k=\tilde{s}\sigma(\tilde{s},\theta)$ where $\sigma$ is the 
differential cross-section. The important consideration for us 
is the behaviour of the various terms in (\ref{B2}) under 
conformal rescaling. We know how to transform $\tilde{s}$ from 
(\ref{s}) and $\tilde{\omega}_p$ from (\ref{vol2}). The Leray 
form $\xi_2$ is conformally-invariant \cite{cb} and so the rescaling 
of (\ref{B1}), at least for the massless case is
\begin{eqnarray}
{\mathcal{L}}_Xf & = & \Omega^2{\mathcal{L}}_{\tilde{X}}f\nonumber\\
&=& \Omega^2C(f,f)\nonumber\\
&=&\int(f_2f_3-f_0f_1) k(\Omega^{-2}s,\theta)\omega_{p^{(1)}}\xi_2.
\label{B3}
\end{eqnarray}
We want the Boltzmann equation (\ref{B3}) to hold up to $\Sigma$ and
so we need to see if it gives another Fuchsian condition. This will
depend on the behaviour of the collision term and specifically on the
behaviour of the kernel $k$ as $\Omega$ goes to zero or equivalently
on the behaviour of $k(\tilde{s},\theta)$ as $\tilde{s}$ increases
without bound. 

If $k$ is sufficiently well-behaved that the integral
in (\ref{B3}) exists in the limit for every $f$ then we get no new Fuchsian
condition. The situation is just like the Vlasov case dealt with in
\cite{a} and we should expect existence and uniqueness with $f_0$ as 
the free data (subject to the Fuchsian constraint corresponding to (\ref{f2}), 
that $\chi_i=0$). The other Fuchsian constraints determine the first
and second fundamental forms of $\Sigma$ in terms of $f_0$. 

However
if $k$ diverges for large $\tilde{s}$ then we do get a new Fuchsian
condition: for the collision term to be finite, the term $(f_2f_3-f_0f_1)$
in the integral must vanish at $\Sigma$. As is familiar from the
literature (see e.g. \cite{eh}), this condition requires the
initial distribution function to take the form
\begin{equation}
 f_0(x^i,p_j) = \exp(-\alpha(x^i)-\beta^a(x^i)p_a)
\label{eqm1}
\end{equation}
for a function $\alpha(x^i)$ and a vector-field $\beta^a(x^i)$, both
defined at $\Sigma$. Next, 
the Fuchsian condition $\chi_i=0$ when imposed on a function $f_0$ 
of the form of
(\ref{eqm1}) requires that the vector $\beta^a$ be proportional to the
normal $N^a$ to $\Sigma$. Now the initial distribution function takes
the form
\begin{equation}
f_0=\exp(-\alpha(x^i)-\beta(x^i)p_0)
\label{eqm2}
\end{equation}
which is formally the distribution for local equilibrium with $\alpha$ and
$\beta$ related to chemical potential and inverse temperature (for
global equilibrium these would be constants). Note that $\beta$ is not
the physical inverse temperature. This would be
$\tilde{\beta}=\Omega\beta$ with the rescaling conventions in use here,
and this vanishes at $\Sigma$ as one would expect. 

The quantity $p_0$ in
(\ref{eqm2}) is understood as a function of $x^i$ and $p_j$ as in
(\ref{p01}) but now
\[
p_0 = (-h^{ij}p_ip_j)^{1/2}
\]
since at $\Sigma$ we can take $V=1$ (and we are supposing for
simplicity that $m=0$). With this in (\ref{eqm2}), the
Fuchsian constraint corresponding to (\ref{f2}) is satisfied provided
$\alpha$ and $\beta$ are related by
\[\beta^4e^{-\alpha}=256\pi^2.
\]
Finally the third Fuchsian constraint, corresponding to (\ref{f3}),
forces the initial second fundamental form to vanish, because the
third moment $\chi_{ijk}$ of $f_0$ vanishes for a distribution
function of the form of (\ref{eqm2}). Thus the data consist of the
initial metric $h_{ij}$ and a function $\beta(x^i)$ which we can call 
the (unphysical) initial inverse temperature. The matter is initially in local
equilibrium, so that the stress-tensor is the stress-tensor for a
perfect fluid. However, there is no reason to suppose that this form
for $f$ will be preserved by the evolution (\ref{B3}).

In conclusion, we see that, as suggested by Rendall and Anguige
\cite{RA}, the Boltzmann case sits between the
perfect-fluid and Vlasov cases studied in \cite{a} and
\cite{at}, and depending on the behaviour of the differential
cross-section at large energies it will, as regards data at the 
singularity, approach the Vlasov case or the perfect fluid case but 
with an extra
function.

\section*{Acknowledgement}
Part of this work was done at the Albert Einstein Institute in Golm in
September 2000 and I am grateful to them for hospitality and 
to Alan Rendall for useful discussions.


\end{document}